\documentclass{article}

\usepackage{kotex} 
\usepackage{arxiv}
\usepackage{array}
\usepackage[utf8]{inputenc} 
\usepackage[T1]{fontenc}    
\usepackage{hyperref}       
\usepackage{url}            
\usepackage{booktabs}       
\usepackage{amsfonts}       
\usepackage{nicefrac}       
\usepackage{lipsum}
\usepackage{graphicx}
\usepackage{booktabs}
\usepackage{multirow}
\graphicspath{ {./images/} }
\usepackage{amsmath} 
\usepackage{hyperref}
\usepackage{placeins}
\usepackage{float}
\usepackage[verbose=silent]{microtype}     

\title{Measuring Falseness in News Articles based on Concealment and Overstatement}

\author{
 Jiyoung Lee \\
  Underwood International College\\
  Yonsei University\\
  Seoul, Republic of Korea\\
  \texttt{jylee1@yonsei.ac.kr} \\
   \And
 Keeheon Lee*\\
  Underwood International College\\
  Yonsei University\\
  Seoul, Republic of Korea\\
  \texttt{keeheon@yonsei.ac.kr} \\
}

\begin{document}
\maketitle
\begin{abstract}
This research investigates the extent of misinformation in certain journalistic articles by introducing a novel measurement tool to assess the degrees of falsity. It aims to measure misinformation using two metrics  (concealment and overstatement) to explore how information is interpreted as false. This should help examine how articles containing partly true and partly false information can potentially harm readers, as they are more challenging to identify than completely fabricated information. In this study, the full story provided by the fact-checking website serves as a standardized source of information for comparing differences between fake and real news. The result suggests that false news has greater concealment and overstatement, due to longer and more complex new stories being shortened and ambiguously phrased. While there are no major distinctions among categories of politics science and civics, it demonstrates that misinformation lacks crucial details while simultaneously containing more redundant words. Hence, news articles containing partial falsity, categorized as misinformation, can deceive inattentive readers who lack background knowledge. Hopefully, this approach instigates future fact-checkers, journalists, and the readers to secure high quality articles for a resilient information environment.
\end{abstract}

\section{Introduction}
In the era of Post-truth, the proliferation of fake news has inflicted negative impacts on individuals and society \cite{Higgins2021}. A notable example is the 2016 U.S presidential election, which highlighted the role of social media in spreading fake news among the public \cite{Allcott2017}. Another salient example is the outbreak of COVID-19 which hindered individuals’ ability to discern truth from falsity. Throughout the pandemic, people turned to unproven drugs and supplements, engaged in hoarding goods or downplayed the danger of disease by taking the matter less seriously. These risky behaviors were all based on non-evidence-based information that has been shared outrageously online, showing the dangers of misinformation that could have life-and-death consequences \cite{Pennycook2020}. Likewise, the proliferation of fake news disrupts the entire news ecosystem by gaining more attention than mainstream authentic news \cite{Shu2017}. 

While recent fake news incidents pay attention to examples of online misinformation, current misinformation research faces several challenges \cite{Zhou2007}. With the lack of agreed upon definition of fake news across different studies \cite{Tandoc2018}, conceptualizing related terms is still in progress. Also, because fake news provides incomplete and noisy data \cite{Shu2017}, a comprehensive and sophisticated review is needed for its detection. Despite the growing concerns of false information harming the quality of journalism; fake news datasets are typically small and lack variation for research purposes. Such deceptive nature which necessitates expertise makes the judgement fake news particularly more difficult. Consequently, fake news research often focuses on individual claims rather than complete news articles \cite{Norregaard2019}, undermining the increasing importance of misinformation caused by journalism.

Hence, with relatively little study on how falsity arise from real news in style and structure, the purpose of this work is to suggest a novel measure to distinguish between false news and real news based on the degrees of falsity. As Reuters Institute Digital News Report\cite{Newman2020} indicates that South Korea has the highest mistrust in journalism among 40 different countries, the Korean articles are explored as a case study.  Given that Korean media typically face fierce competition which leads to lack of fact verification, the result should help explain how false information by journalists undermines trust in journalism. It is important to note that misinformation in Korea is not limited to specific media outlets and persists without correction by other media outlets reproducing it in the same manner. In this context, while the term fake news is used interchangeably with false information \cite{roozenbeek2019fake}, “false news” is consistently used throughout the paper to refer to incorrect information produced by journalists. Furthermore, the work employs both linguistic features and content analysis to aid in identifying the distinctive characteristics between false and real news, thereby alerting readers of potential false news articles. Beyond building automated solutions using machine learning to detect false news, adopting simple metrics to measure falsity can pave new useful ways to tackle malicious online information.

\subsection*{Definition: What is False News?}
\label{sec:headings}
A clear definition of false news is important for the study. This section is an overview of the terminology of false news by introducing its broad definitions, various forms, and the definition studied in this work to consolidate the terminology and scope used throughout the paper. First, false news refers to the fabrication of factual or relevant information that is different to the reality. Given that all false news has low level of facticity, false news can also be divided into two broad categorizations according to the producer’s intention.
\paragraph{Disinformation:}
\underline{deliberate} presentation of (typically) false or misleading claims as news, where the claims are misleading by design \cite{Gelfert2018}
\paragraph{Misinformation: }
\underline{incorrect or misleading} information which undermines the credibility of news outlets \cite{Lazer2018}

There is a wide spectrum of false news that encompasses multiple categories extending beyond the complexities of disinformation and misinformation. The most common term in the field is fake news, which emphasizes  disinformation. For example, the UNESCO divides fake news into three categories of misinformation, disinformation and malinformation \cite{Ireton2018}. Allcott and al\cite{Allcott2017}, includes unintentional mistakes, rumors from unverified resources, conspiracy theories, satire, false statement, and misleading reports in the close cousins of fake news \cite{Allcott2017}. Claire Wardle situates fake news in the larger context of misinformation and disinformation \cite{wardle_fake_news}, with seven types of information disorder from satire and parody, to  manipulated content \cite{Wardle2018}. Although different types of fake news partially overlap one another depending on the scope and interpretation, there are important differences that make the distinctions. The elements of distinction involve (i) facticity, (ii) author’s intention, and (iii) quality of journalism, all of which are rated and organized differently by researchers \cite{Tandoc2018} \cite{Zhou2018} \cite{wardle_fake_news}.

While current definition of fake news focuses on disinformation which emphasizes author’s intention to deceive \cite{Tandoc2018}, this work concentrates on assessing the level of falsity in real world articles. Thus, it employs the term “false news” which aligns more closely with misinformation to make a distinction to “fake news” which is more associated with disinformation. The term false news is used as an umbrella term to describe false articles collected by Korean fact-checking website-SNUFactCheck, which is the biggest fact-checking website that provides fact checked articles. In the context of this study, false news refers to misinformation generated by journalists and confirmed by fact-checking websites. The purpose of using false news is driven by the fact that intentionality is difficult to measure leading to scholars using articles that are proven false \cite{Damstra2021}. Building on prior research which demonstrated that partially untrue messages can spread more widely than truth \cite{Vosoughi2018}, this study focuses on any falsity produced by journalists, whose primary duty is to produce quality information.

Hence the dataset contains misinformation which is initially reported as true, but later found to be false \cite{Lewandowsky2012}, and yet to be corrected. Such misinformation is particularly important because one of the main challenges of misinformation studies was its scarcity \cite{Zhou2007}. Also, as readers commonly encounter reconfigured misinformation rather than completely fabricated fake news \cite{Brennen2020}, false news in the type of misinformation tend to be more pernicious to readers.

\subsection*{Measuring falseness}
Three streams of literature related to measuring false news articles are provided herein. First, in an effort to identify deception using linguistic cues, methods for evaluating deceptive content have been explored. The study by Afroz et al. examines how authors’ linguistic features change when they try to obfuscate their writing style using word based features such as total characters, frequency of words \cite{Afroz2012}. Another linguistic based cues for detecting deception conducted by Zhou et al. proposes to look at the amount of words, noun phrases, sentence complexity, uncertainty, and informality features \cite{Zhou2004}.

Second, in the NLP related areas, false news identification often involves a combination of different quantifiable metrics and machine learning techniques for its prediction \cite{Rubin2016} \cite{Monteiro2018}. An important finding by Horne et al. compares fake to real news along with satire news using content analysis and linguistic features \cite{Horne2017}. The research presents that fake news articles uses simpler, repetitive content in the text body with significantly more lexical redundancy than real news articles. The Support Vector Machine model was used to test the predictive power of the extracted linguistic features. Similarly, Silva et al introduces a Portuguese news dataset, corpus of aligned true and fake news in different categories, to analyze linguistic-based features, and to detect fake news using machine learning methods \cite{Silva2020}. 

Lastly, with the growing interest in assessing the information quality, factual density is often used to measure the informativeness of the text. By assessing the quality of content using fact count in an article normalized by the text size, Lex et al. identifies featured to non-featured articles in Wikipedia \cite{Lex2012}. The factual density measure is often compared with other measurements like simple metric – word count – to estimate the article quality. For example, Blumenstock extracted readability metrics, syntactic features, structural features and part of speech tags and tested simple to complex classification schemes \cite{Blumenstock2008a}. The result indicates that such simple methodology of using word counts and article length over complex methods can be a good predictor of deciding featured articles \cite{Blumenstock2008b}.

Despite these efforts in news articles measurement, yet no study has analyzed the ratio of true to untrue information to determine when the content becomes false \cite{Egelhofer2019}. Further, many experiments did not exclusively use news articles of the same topic. Thus, this work builds on these streams of literature by analyzing the proportion of information of aligned false and real news with similar topics, to examine its falseness and fill this gap in research. The informativeness of news articles is related to frequency count and document length\cite{Blumenstock2008b}. Based on the information manipulation theory, violation of the principles of quantity, quality, manner and relevance of information results in misinformation \cite{McCornack1992}, and therefore the degree of falseness can be measured in terms of the sufficiency in word quantity. Likewise, as described by Zhou et al., misinformation can be divided into four types, mainly concealment, ambivalence, distortion and falsification, whereby concealment is a violation of appropriate quantity of information \cite{Zhou2007}. Hence, given that news articles from the same source of information provided by the fact-checking website can vary among each other in terms of its quantity and quality, concealment and overstatement are used as two indicators of falseness.

\hspace{1cm}\textbf{Concealment:} hiding important information 
 
\hspace{1cm}\textbf{Overstatement:} overwriting information 

That is, as facts are distorted by manipulation of information, news articles based on concealment refers to news contents that are insufficiently written to mislead the readers. On the other hand, news based on overstatement refers to news contents that often add in exaggerated or inflammatory words to attract readers \cite{Au2021}. Both of these indicators of false news, concealment and overstatement, are used to reveal the characteristics of falsity in news. From the research conducted until now, as false news is more descriptive; use fewer technical words with more lexical redundancy and may differ based on their topics, this study hypothesizes the following.

 \textbf{H1:} false news has greater overstatement than real news due to intended word redundancy to exaggerate, and mislead the readers.

\textbf{H2:} politics category contains greater falsity than any other categories of false information as they are more sensational with emotional statements of politicians

The political news is predicted to make the most significant difference as they are more viral than other categories \cite{Vosoughi2018}. Finally, many news consumers are worried about online misinformation because it is more difficult to define than completely made-up news articles. This should help understand how misinformation produced by journalists, whose primary duty is to deliver truth rather than captivating stories, encourage readers to unconsciously accept partially false information. As hardly any news article is entirely false or real, and there is yet an ongoing disagreement on conceptualizing false  news \cite{Molina2019}, providing a simple  novel measure in journalistic articles should help filter out misinformation at an early stage.

\section{METHODOLOGY}
Collecting false information is an arduous task. Although there are possible fake news datasets such as the BuzzFeed Fake News Dataset \cite{potthast-etal-2018-stylometric} and the LIAR dataset \cite{Wang2017}, they are not in use in this experiment because they comprises short posts, statements or articles fabricated primarily for satirical purposes by fake news websites. As this work aims to study the falsity in real articles of journalistic format, these datasets were deemed unsuitable. Second, another possibility could involve crowdsourcing false news for comparison with legitimate news \cite{perez-rosas-etal-2018-automatic}. However, such news articles would differ from naturally occurring misinformation found online. Therefore, this study focuses on articles produced by journalists that inadvertently undermine trust in journalism. Lastly, even fact-checking websites often simply present commentary on news or rumors without URLs to misinformation articles, reducing the size of dataset. As the alignment of real news and false news is needed, a new dataset had to be newly created. 

\subsection*{Data Collection: Fact checking website}
Defining false information has always been difficult because they are partially true or partially false at the same time. Thus, the possibility of verification was the focus in this work to determine the falsity of information \cite{SHAHI2021100104}.  Also, there is difficulty in finding false news as they are often deleted. Hence this study used fact-checking website (Factcheck snu) – a prominent Korean fact checking website - which provides URLs to misinformation articles that are still available online (Table.1.). 

The fact checking website provides ways to detect false news articles, with explanations on how they rate their Accuracy. It also provides URLs for both false and real news articles relevant to the topic under discussion. Despite the methodological criticism of many fact checking news websites, the dataset collected offer meaningful and transparent journalistic perspective \cite{Nieminen2021},. It addresses the limitations of using artificially manipulated false news articles by sourcing genuine articles from various news outlets to better reflect the reality. This guarantees the validity of the articles, as all articles are sourced by fact-checkers either through archived posts or referenced links to news website.

\begin{table}[ht]
\centering
\caption{Example Data from Fact-Checking Websites}
\begin{tabular}{|>{\raggedright\arraybackslash}p{2cm}|>{\raggedright\arraybackslash}p{5cm}|>{\raggedright\arraybackslash}p{5cm}|}
\hline
\textbf{Source} & \textbf{Factcheck org} & \textbf{Factcheck snu} \\
\hline
Cited / archived misinformation & \vspace{0pt}\includegraphics[width=5cm, height=4cm]{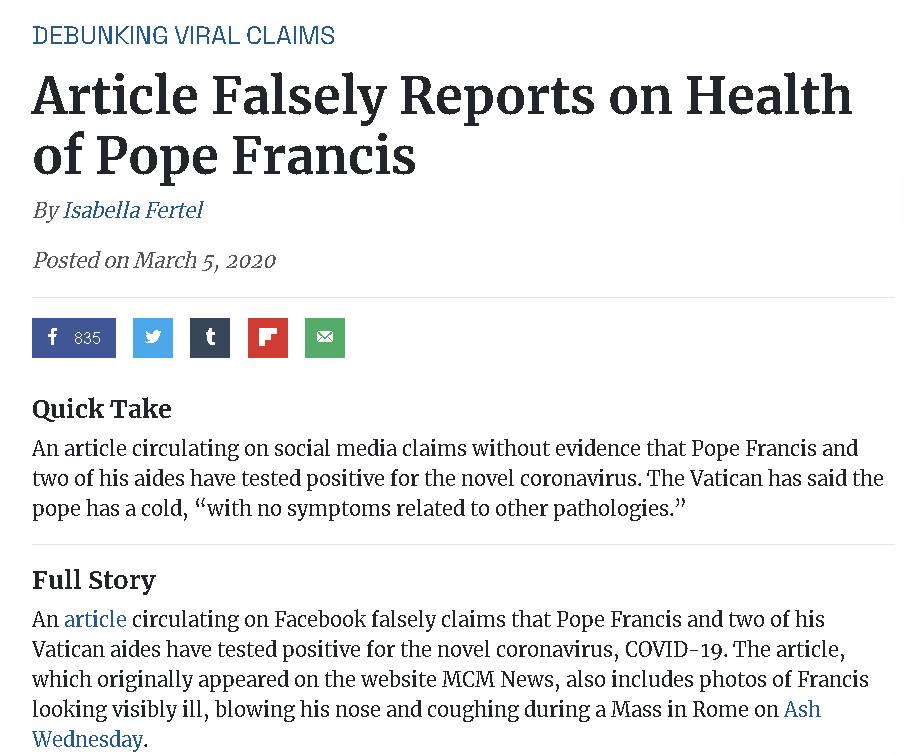} & \vspace{0pt}\includegraphics[width=4.5cm, height=4cm]{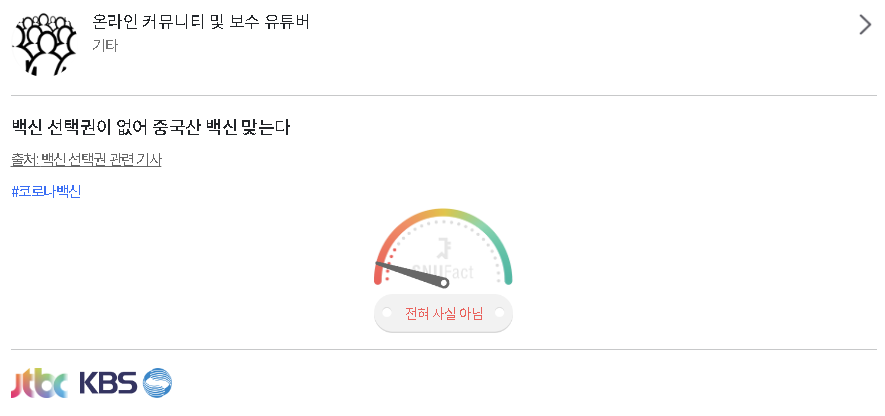} \\
\hline
Webpage & \vspace{0pt}Webpage \includegraphics[width=5cm, height=4cm]{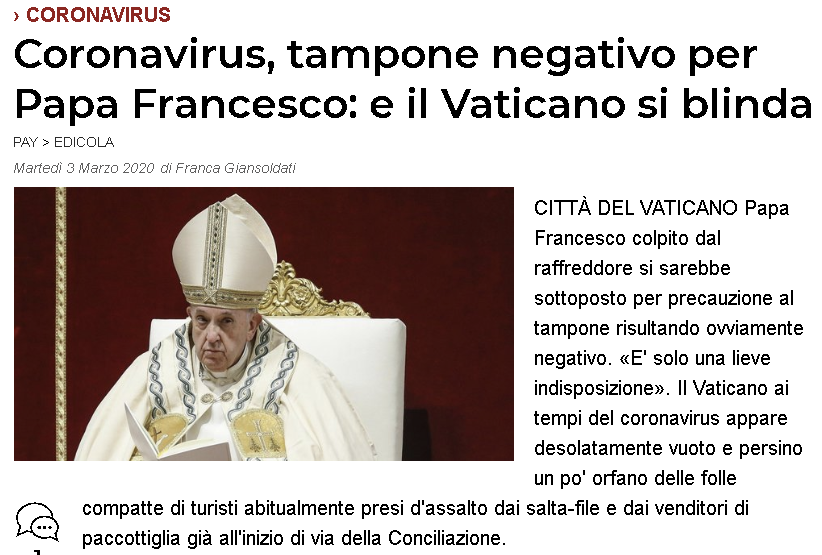} & \vspace{0pt}\includegraphics[width=5cm, height=4cm]{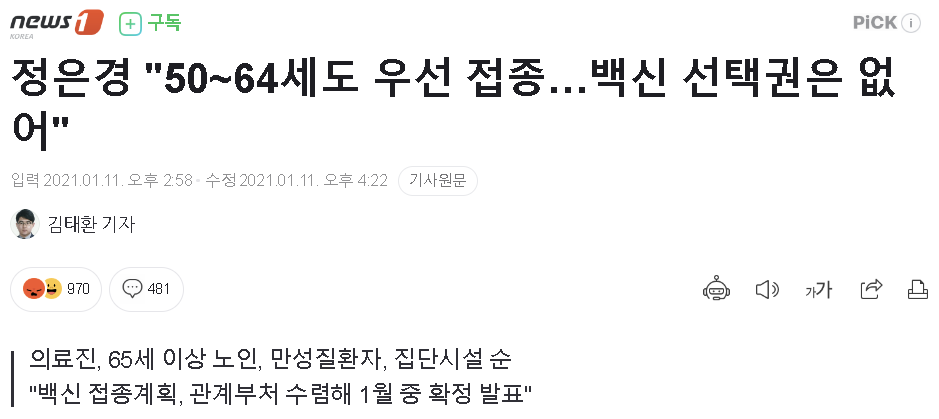} \\
\hline
Judgement & \vspace{0pt}Viral claim & \vspace{0pt}False \\
\hline
link & \vspace{0pt}\url{https://factcheck.snu.ac.kr/facts/show?id=2739} & \vspace{0pt}\url{https://factcheck.snu.ac.kr/facts/show?id=2739} \\
\hline

\end{tabular}
\label{tab:example}
\end{table}

\subsection*{Data explanation}
Hence the data set consists of false news matched with real news, and the full story written by the fact-checking website on the same issue. 

The full story acts as a standardized information source, which interprets data using nonpartisan sources. This neutral source not only addresses the facts of the same issue but also includes arguments on how they reached their conclusions. Using the full story should illuminate the lexical changes of primary information into actual reporting and the emergence of misinformation when compared with false and real news.

Along with the full story, SNU fact checking website provides URLs to both false and real news articles that are used in this experiment. These articles are collected coherently in the way that is agreed upon by many fact checkers. To lessen the subjectivity of false news even further, this study used false news judged as “False” within their six different rating scales (Not fact at all(false), mostly not fact(false), half fact, mostly fact, and fact).

By using the articles collected by the fact-checking website, the study collected 20 articles from science category, 34 articles from politics category, and 32 articles from civics category. These three categories serve as criteria by which many fact-checking websites classify articles. Other categories include economics, culture, and international but some articles in these categories were also in part of the three main categories. In total, the dataset add up to 43 false news and 43 real news articles. Many of the topics concern famous issues related to the COVID-19 pandemic, which is the period of uncertainty with various types of misinformation about many different topics \cite{Brennen2020}. Table 1. shows example data from the science category.

\begin{table}[ht]
\centering
\caption{Example Data from the Science category}
\begin{tabular}{|>{\centering\arraybackslash}p{3cm}|>{\centering\arraybackslash}p{3.5cm}|>{\centering\arraybackslash}p{3.5cm}|>{\centering\arraybackslash}p{3.5cm}|}
\hline
\textbf{} & \textbf{Fact-checked article} & \textbf{Full story} & \textbf{False information} \\
\hline
\textbf{Date} & 24.08.2020 & 26.08.2020 & 25.08.2020 \\
\hline
\textbf{Content} & 실내 마스크 착용이 비과학적일 뿐만 아니라 건강에 해악을 초래한다고 주장했다. 교육 현장의 차별로도 이어진다는 주장도 내놨다. 뉴스톱이 팩트체크했다. & 이 중 “마스크의 장기 착용은 밀폐된 좁은 공간에서 장기간 숨 쉬는 것과 같아 건강을 해친다”는 항목이 있다.  그러나 어떤 통계나 근거도 제시하지 않는다. & 마스크 착용에 대한 부작용이 상당하다고도 했다. 또 만성적인 저산소증에 시달리는 등 부작용이 많다”고 설명했다. \\
\hline
\textbf{Source} & News website & Fact checking website& News website  \\
\hline
\textbf{Definition} & \multicolumn{3}{>{\centering\arraybackslash}p{10.5cm}|}{Misinformation on wearing masks indoor to be the cause of hypoxia. } \\
\hline
\end{tabular}
\end{table}

According to Silva et al., there are insufficient false news datasets that are matched together with real news \cite{Silva2020}. Also, to the best of our knowledge, there is yet no false news dataset that contains false news, real news and the full story written by the fact-checking website. Thus, collecting new dataset with aligned false and real news articles in different categories that arise based on the same full story is another contribution of this work. 

To compare linguistic features of real and false news contents, lexical diversities were often analyzed using Type-token ratio. As shown in Table.2, real news tends to be larger in text size than falsee news, in number of tokens, nouns, and sentences. while type-token ratio, which shows the lexical variation, of real news tends to be smaller than false news. This finding is similar to Horne et al. and Pérez-Rosas et al that use English and Portuguese fake news dataset in their analysis \cite{Horne2017} \cite{perez-rosas-etal-2018-automatic}.

\begin{table}[ht]
\centering
\caption{Basic Analysis of False and Real News Dataset}

\begin{center}
\begin{tabular}{|c|c|c|c|}
\hline
 & \textbf{Features} & \textbf{False News} & \textbf{Real News} \\
\hline
\multirow{4}{*}{\textbf{Contents}} 
 & Total number of tokens & 14,555 & 23,726 \\
\cline{2-4}
 & Total number of POS & 13,869 & 23,471 \\
\cline{2-4}
 & Total number of sentences & 889 & 1335 \\
\cline{2-4}
 & Type-token ratio(TTR) & 0.595 & 0.501 \\
\hline
\end{tabular}
\end{center}
\end{table}

\subsection*{Falseness Measurement}

The contents of the articles are only used for the experiment. Before analyzing the content, all unnecessary contents such as correction notes, pictures, captions, date, and names of the author are to be removed. Then, natural language processing (NLP) for Korean language, Mecab in KoNLPy is used herein to handle Korean news articles, to helps understand the difference in stylistic features, and in analyzing the part of speech (POS) elements of false and real news articles.
To measure the falseness of news articles based on concealment and overstatement, the assumption is that all consequent articles are based on one specific full story. The full story functions as the main source to compare false and real news. Hence, this should help answer the proportion of information that is either lost or added compared to one specific full story. Assuming that nouns are representative features of information in each article, the difference in falseness measurement is based on the size of nouns that is either added or lost in false or real news articles and are calculated as in table 4.

\begin{table}[ht]
\centering
\caption{Example Calculation of Falseness}
\begin{center}
\begin{tabular}{|c|c|c|c|}
\hline
\textbf{Indicator} & \textbf{Full story ($\mathbf{T_1}$)} & \textbf{False News ($\mathbf{T_2}$)} & \textbf{Falseness} \\
\hline
Concealment & 
\begin{tabular}{c}
$[\text{"살균", "소독제", "폐",}$ \\ $\text{"질환", "예방"}] = 5$
\end{tabular} & 
\begin{tabular}{c}
$[\text{"소독제", "폐",}$ \\ $\text{"질환", "유발"}] = 4$
\end{tabular} & 
\begin{tabular}{c}
$\frac{|T_1 \cap T_2|}{|T_1|} = \frac{5 \cap 3}{5} = 0.4$
\end{tabular} \\
\hline
Overstatement & 
\begin{tabular}{c}
$[\text{"살균", "소독제", "폐",}$ \\ $\text{"질환", "예방", "사용"}] = 6$
\end{tabular} & 
\begin{tabular}{c}
$[\text{"소독제", "폐",}$ \\ $\text{"질환", "유발"}] = 4$
\end{tabular} & 
\begin{tabular}{c}
$\frac{|T_2 \cap T_2|}{|T_2|} = \frac{4 \cap 3}{4} = 0.25$
\end{tabular} \\
\hline
\end{tabular}
\end{center}
\end{table}

\section{Results}

\subsection*{Falsity Measurement: Simple Linear Regression}

\begin{figure}[htbp]
    \centering
    \caption{Simple Linear Regression for False News and Real News}
    \includegraphics[width=\textwidth]{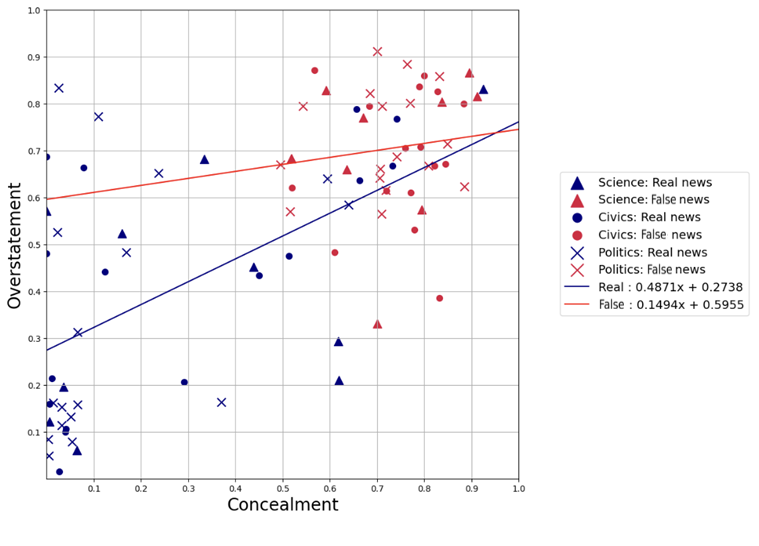}
\end{figure}

For the purpose of comparison, the line of best fit $y = \beta_0 + \beta_1 x$ $+ \epsilon$ is regressed for both real and false news. As for the real news, the line of best fit is y =0.4871x+0.2738, and as for the false news the line of best fit is y =0.1494x+0.5955. While R-squared evaluates how close the data points are to the line of best fit, the R-squared is 0.2624 for real news, and 0.0171 for false news. The result indicates that false news has greater concealment and overstatement, due to longer and more complex new stories being over simplified and exaggerated. That is, by comparing both false and real news to the full story, more information is lost and or added to become false. Also, given the slope of false and real news, false news tends to have greater overstatement, adding relatively more information than real news, while real news tends to have greater concealment, with more concise sentences. Along with the result of Table3., Figure 1. suggests that false news has greater word diversity (TTR), because they have greater overstatement than real news. This shows that if journalists fail to write articles properly and succinctly, there is a potential risk for the news to become false.

\subsection*{Statistical Tests: Comparing the Slopes Using T-test}

This section tests whether the slopes for both false and real news are equal using the t-test which is appropriate for small sample size. The null hypothesis proposes H0:$\beta_1$ =$\beta_2$ i.e. $\beta_1$ - $\beta_2$ = 0 that there is no significant difference between the two slopes, and the alternative hypothesis proposes H1:  $\beta_1$ ≠ $\beta_2$i.e. $\beta_1$  – $\beta_2$ ≠ 0 that there is a significant difference between the two slopes. Given that the slope for real news is 0.4871 and the slope for false news is 0.1494, the t-test value is 1.425. The p-value for the test statistics with degrees of freedom ($n_{1}$ + $n_{2}$ – 4) is 4.539e-24, which is smaller than alpha 0.05. Hence, it can be concluded that the null hypothesis is not rejected, and a significant difference is found between the slopes of false and real news.

\begin{table}[ht]
\centering
\caption{Overall t-test result}
\begin{tabular}{|c|c|c|}
\hline
t-value & Degrees of Freedom & P-value \\
\hline
14.1547 & 85 & 4.539e-24 \\
\hline
\end{tabular}
\end{table}

\subsection*{Statistical Tests: Comparing the Slopes Using Mann Whitney U-test}

Additional statistical analysis was conducted using the Mann-Whitney U test, which is a nonparametric test of two random populations with a small sample of population with an unknown normality. Instead of using the mean as in the t-test, Mann Whitney U-test uses the median to compare the two independent samples, and measures the ranks to examine both the location and shape. Given the characteristics of the dataset, this helps better understand the difference between false and real news, and examine whether one variable tends to have values higher than the other. To do so, this time the test was conducted for each variable, concealment and overstatement of false and real news samples. The null hypothesis is that there is a no difference between the false and real news, and the alternative hypothesis is that there is a difference between the two (that is, they are not equally distributed).

When comparing the concealment of false and real news at a significance level of 0.05 with 2-tailed hypothesis, the result is significant at p<0.05 (p-value is 1.523e-11). And, when comparing the overstatement of false and real news at a significance level of 0.05 with 2-tailed hypothesis, the result is also significant at p<0.05 (the p-value is 3.945e-8). Hence, it can be concluded that the concealment is a better indicator than overstatement to discern between false and real news, as false news tends to hide information by writing articles imprecisely compared to real news. Thus, articles lacking in details than adding more details should be read more cautiously as they are likely to be false

\begin{table}[ht]
\centering
\caption{Overall Mann-Whitney result}
\begin{center}
\begin{tabular}{|c|c|c|c|}
\hline
Indicator & Z-score & P-value & Result (p<0.05)\\
\hline
Concealment & 6.7457 & 1.523e-11&significant \\
\hline
Overstatement & 5.4933& 3.945e-8 &significant \\
\hline
\end{tabular}
\end{center}
\end{table}

\subsection*{Falseness Measurement: Decision Boundary}

The difference between false and real news content is further explored using decision boundaries in classification models. The classification methods include logistic regression (LR), random forest (RF), quadratic discriminant analysis (QDA), Naïve Bayes (NB), support vector machine (SVM) (Silva et al. 2020), and decision tree (DT) from MLxtend. It provides machine learning tools for data analysis, which uses supervised learning models to find an optimal classification. The results in Figure 2 shows that the upper blue region denotes the data of class “Real”, and that the lower red region denotes the data of class “False”. Along with the findings in Figure 1, with higher proportion of overstatement of real news to false news, it is classified in the upper region of the decision boundary. In contrast, with higher proportion of concealment of false news to real news, it is classified in the lower region of the decision boundary.

\FloatBarrier
\begin{figure}[H]
    \centering
    \caption{Decision boundaries}
    \includegraphics[width=\textwidth]{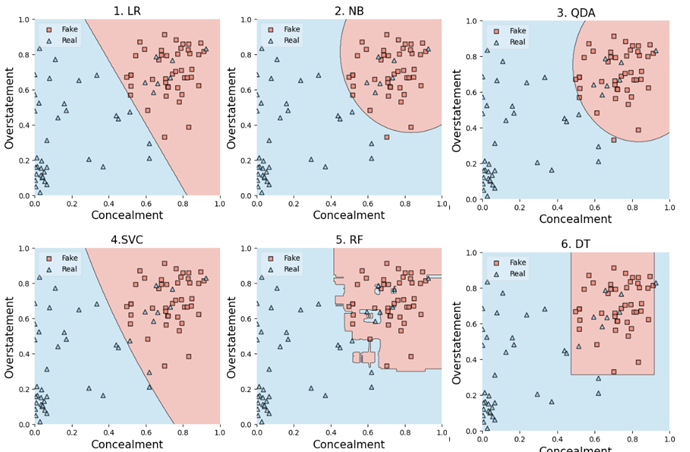}
\end{figure}

Table 7. shows the order of accuracy. While the Logistic Regression, Naïve Bayes and Quadratic Discriminant Analysis (QDA) shows the highest classification accuracy of 0.92 in discriminating false to real news, the Decision Tree model performs low accuracy of 0.87. Hence, along with Figure 2, the result indicates that effective decision boundary must be linear.

\begin{table}[ht]
\centering
\caption{ Classification Accuracy of Decision Boundaries}
\begin{center}
\begin{tabular}{|c|c|>{\centering\arraybackslash}p{3cm}|>{\centering\arraybackslash}c|}
\hline
No. & Classifier & Accuracy &Standard Deviation \\
\hline
1 & Logistic Regression& 0.92 & 0.03\\
\hline
2 & Naïve Bayes & 0.90&0.02\\
\hline
3 & Quadratic Discriminant Analysis & 0.90 &0.06 \\
\hline
4 & Support Vector Machine& 0.88& 0.02 \\
\hline
5 & Random Forest& 0.87 &0.02 \\
\hline
6 & Decision Tree& 0.87 &0.02 \\
\hline
\end{tabular}
\end{center}
\end{table}

\subsection*{Falseness Measurement: Difference by Categories}

\FloatBarrier
\begin{figure}[H]
    \centering
    \caption{Simple Linear Regression by Categories}
    \includegraphics[width=\textwidth]{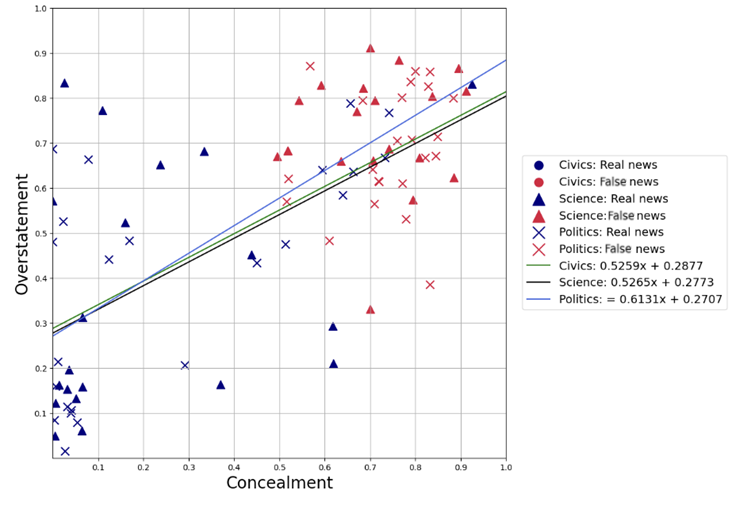}
\end{figure}

The simple linear regression by different categories indicates that the line of best fit for the civics category is -0.5259x+0.2877 with R-squared of 0.493; the line of best fit for the science category is 0.5265x+02773 with R-squared of 0.4018; and the line of best fit for the politics category is 0.6131x+0.2707 with R-squared of 0.525.

Since the ratio of overstatement and concealment may vary depending on the topic, linear regression analysis is used separately to test if the falseness differs by categories. As for the difference by category of false and real news, politics category tends to have higher proportion of overstatement compared to concealment of information. Considering that political articles often rely on rapidly emerging breaking news that is often exaggerated, this seems reasonable. As for the science category and civics category, false news articles tend to have both greater overstatement and concealment of information than real news. Given the nature of complex scientific findings and the purpose of science articles to accurately deliver details of scientific studies, false news tend to lack such characteristics. As explained by Lewandowsky et al, high concealment of information in scientific reporting may be problematic, because oversimplification in scientific results risks in misunderstanding and misrepresentation \cite{Lewandowsky2012}. 

\FloatBarrier
\begin{figure}[htbp]
    \centering
    \caption{Confidence Ellipses in a Scatterplot with Different Categories}
    \includegraphics[width=0.9\textwidth]{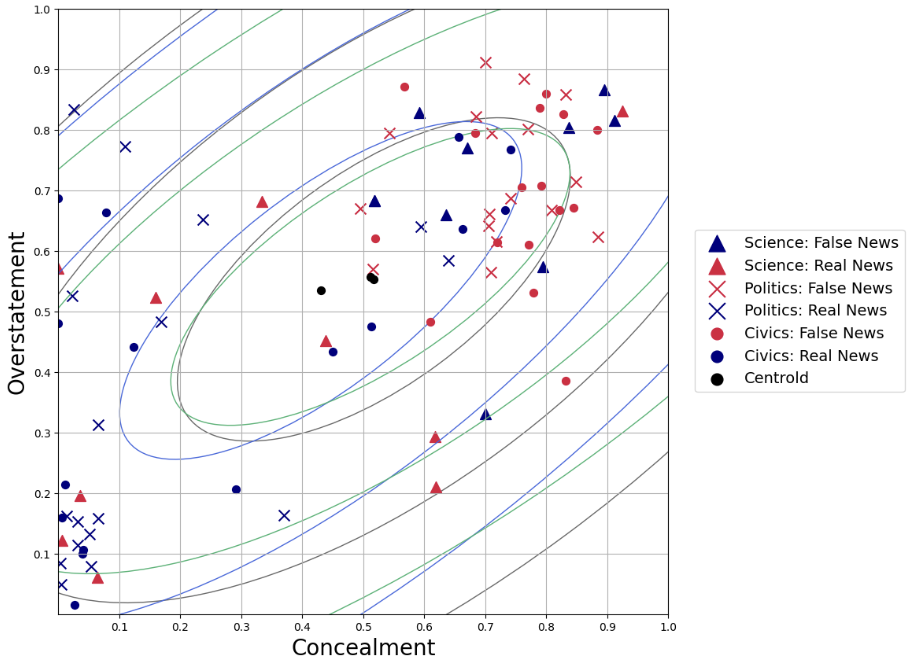} 
\end{figure}

\FloatBarrier
\begin{table}[htbp]
\centering
\caption{Confidence Ellipse and Distance for False and Real News Categories}
\begin{tabular}{|p{5cm}|p{2cm}|p{2cm}|p{2cm}|} 
\hline
\textbf{Category} & \textbf{Civics} & \textbf{Science} & \textbf{Politics} \\
\hline
Centroid (Concealment, Overstatement) & 0.512, 0.557 & 0.517, 0.553 & 0.438, 0.534 \\
\hline
Mahalanobis Distance & 1.981 & 1.920 & 2.157 \\
\hline
\end{tabular}
\end{table}

In figure 4, the confidence ellipses are provided to add a visual summary to a scatter plot to examine its dispersion. The result represents the three standard deviations for each category, and the respective centroids. The shapes of the confidence ellipse also show how strongly the data points are correlated. By visual examination, all three categories are widely dispersed from its centroid that represents the mean point which obscure the distinction among the three. As for the Mahalanobis Distance, politics articles have slightly greater Malalanobis distance than other doamins, followed by civics and science.

\subsection*{Linguistic Features: Part of Speech Comparison}
Assuming that part of speech (POS) tagging has a predictive power, this section analyzes the part of speech (POS) tagging to distinguish between false to real news. Hence, the linguistic features using the part of speech taggers from NNG (common noun), NNP (proper noun), NP (pronoun), VV (verb), VA (adjective), MAG (adverb), SL (foreign language), to SN (number) are provided. Overall, false news has more linguistic features in concealment and overstatement in almost all taggers. Moreover, false news is more likely to add numbers than real news; however, this should be further studied with greater dataset.

\FloatBarrier
\begin{table}[H]
\centering
\caption{False and Real News Concealment}
\begin{tabular}{|p{2cm}|p{2cm}|p{2.5cm}|p{2.5cm}|p{2.5cm}|}
\hline
\multirow{2}{*}{\textbf{Feature}} & \multirow{2}{*}{\textbf{Total}} & \multicolumn{3}{c|}{\textbf{Category}} \\ 
\cline{3-5}
 & & Science & Civics & Politics \\
\hline
NNG & false > real & 1,017 > 515 & 1,589 > 524 & 1,917 > 309 \\
\hline
NNP & false > real & 69 > 36 & 114 > 29 & 186 > 23 \\
\hline
NP & false > real & 12 > 8 & 18 > 5 & 19 > 3 \\
\hline
VV & false > real & 40 > 21 & 42 > 17 & 50 > 12 \\
\hline
VA & false > real & 30 > 15 & 38 > 13 & 36 > 6 \\
\hline
MAG & false > real & 92 > 46 & 128 > 39 & 131 > 14 \\
\hline
SL & false > real & 68 > 53 & 64 > 22 & 82 > 13 \\
\hline
SN & false > real & 104 > 57 & 307 > 113 & 277 > 53 \\
\hline
\end{tabular}
\end{table}

\begin{table}[htbp]
\centering
\caption{False and Real News Overstatement}
\begin{tabular}{|p{2cm}|p{2cm}|p{2.5cm}|p{2.5cm}|p{2.5cm}|}
\hline
\multirow{2}{*}{\textbf{Feature}} & \multirow{2}{*}{\textbf{Total}} & \multicolumn{3}{c|}{\textbf{Category}} \\ 
\cline{3-5}
 & & Science & Civics & Politics \\
\hline
NNG & false > real & 876 > 570 & 1256 > 1142 & 2000 > 1006 \\
\hline
NNP & false > real & 85 > 43 & 102 > 81 & 166 > 110 \\
\hline
NP & false > real & 10 < 13 & 20 = 20 & 40 > 30\\
\hline
VV & false > real & 15 < 20 & 51 >43 & 48 > 24 \\
\hline
VA & false > real & 15 < 19  & 27 < 28 & 39 > 31 \\
\hline
MAG & false > real & 38 < 58 & 96 > 94 & 163 > 101 \\
\hline
SL & false > real & 24 > 20 & 18 > 10 & 54 > 25 \\
\hline
SN & false > real & 113 > 29 & 108 > 67 & 118 > 51 \\
\hline
\end{tabular}
\end{table}

\section{DISCUSSION}
The purpose of this work is to assess the degree of falsity contained in news articles by measuring the level of concealment and overstatement. Through these two criteria, it aims to understand how journalists can distort truth into misinformation by presenting it in different writing styles. The main contributions of this study are: reorganizing the term false news, providing a new dataset which juxtaposes false information and real new alongside the full story written by fact-checking websites, and addressing the degree of falsity. While a more sophisticated framework in defining false news is necessary, this novel approach should help understand the challenges posed by partially true or false information and its potential harm to society. 

Regarding the first hypothesis that evaluates the degree of falsity, the result indicates that false news contains relatively more redundant information than real news in covering the same issue. Additionally, false news tends to omit important information more frequently than real news. Using the line of best fit, false news adds in more information with greater overstatement and loses more information than real news with greater concealment. According to the statistical analysis, the two line of best fit showed that there is a significant difference between false and real news, and that both concealment and overstatement indicators were useful in determining their difference. While false news articles are shorter than real news as can be found in table 3., they have more lexical diversity (TTR). When POS taggers are analyzed, false news had greater number of different taggers from words that are concealed and overstated. With regards to the second hypothesis that examines news articles from different categories, there are subtle variations in the degree of overstatement and concealment between false and real news. While they all show that false news has greater overstatement and concealment, it is challenging to assess their differences due to a similar slope in the line of best fit.

\section{LIMITATION}
One of the main challenges of false news detection is the lack of an agreed definition \cite{Zhou2018} \cite{Clayton2019} \cite{Tandoc2018}. Given the complexity and diversity of false news with its varying features such as the content, headlines, presentation, and tone of argument, fact checking becomes subjective and unclear. This is why many fact checking websites rate the truthfulness of news articles rather than simply categorizing them as true or false \cite{Wang2017}. Certain news articles that have misleading headlines and some correct content require level of expertise to understand, which makes it difficult of readers to discern. Given that evaluation of falsity is critical in false news study, extra effort is necessary to ensure the validity and completeness of false news datasets \cite{OBrien2018}. 

Another limitation is the lack of non-noisy false news dataset for the studies  \cite{OBrien2018} \cite{Wang2017} \cite{Rubin2016}. Although there are fake news open datasets, they are often crafted, lack variety, incomplete, or do not align vis-à-vis legitimate ones. The main adversity in this work is also finding suitable and reliable data for the experiment. While misinformation arising from journalism causes severe infodemic, in most cases, they are corrected before being collected. Likewise, it was particularly difficult to find false news in journalistic format because false information often originate from rumors, posts, works of fiction, government and many more \cite{Lewandowsky2012}. With relatively small dataset that is manually collected, larger false news dataset with more diversified topics is essential to facilitate the studies in false news \cite{TorabiAsr2019}.

Furthermore, in the context of the Korean journalistic environment, there is yet no renowned websites created to deliberately spread fake news for satirical purposes. Hence, the term false news has lost connection among manipulated or misleading contents, blurring its definition by obsessing with the phrase “fake news”. Consequently, the focus is articles with misleading contents which is prevalent online because they are easily reproduced by journalists or news consumers, and contribute to the perpetration of disinformation. However, finding a suitable misinformation was also challenging because of its nature. For instance, many real news articles were copies of the original full story written by the fact-checking websites, and therefore may have affected the length of articles. Given that misinformation can be altered or fixed over time, fake news in disinformation format can be used for future analysis. In this case, aligning same topic with similar stories would be needed to replicate the experiment.

\section{CONCLUSION}
Hence, the findings imply the dangers of journalistic laziness in covering the same source of information. Many breaking news updates were easily replicated and reproduced without undergoing thorough fact verification. This is due to excessive competition among breaking news coverage, occurring regardless of media outlets. Like the example of misrepresentation by media of global extinction published in Nature \cite{Ladle2005} , the inability of journalists to fully capture the full story risks in oversimplification or slips in word choice that lead to misreporting. From the finding of this work, misinformation is more prone to occur from both loss of information or simplification, and addition of information or hyperbole. This tendency of concealment by false news becomes more problematic as people often prefer simple explanations over complex explanations \cite{Johnson2019}. The mainstream news organizations, which are perceived to provide reliable information, also reproduce misleading content by recopying behavior, causing serious consequences \cite{Matusitz2007}. Once misinformation iterates in social media, repeated exposure to the same false statement can lead to its acceptance and solidify belief \cite{Begg1992} \cite{Hasher1977}. Hence, given the high credibility in journalism than rumors or claims on social networks, misinformation reproduced by journalists is more dangerous because readers are less sensitive to assessing its message in sufficient detail. Thus, future research should explore ways to evaluate journalists’ ethical practices to lessen journalistic laziness in their news reporting, and understand its consequences.

To maintain the flow of high quality information in the future, effort from journalists, readers, as well as the fact-checkers is needed. First, the journalists should not be too reliant on one single source of information or secondhand press reports and should cross-check facts with other sources of information. As concealed information is a characteristic of false news, detailed explanations with tentative language about an issue can be a strategy to prevent its risk. If journalists find problems in their articles, they should behave with integrity by acknowledging mistakes in their article, and announce that corrections have been made to the later version. Second, the readers must make a habit of going upstream to the original story to evaluate the news article, and increase individual skepticism over information. Providing nudges to news consumers, encouraging them to think analytically about potential misinformation can also reduce the sharing of false news \cite{Andi2020}. Lastly, effective communication between press release and news coverage is essential \cite{Lee2016}. The primary source of information should ensure clear and unambiguous messaging, so that journalists that use it can accurately convey the same message without distortion. Each information providers should understand how their coverage is transformed and reinterpreted into news articles, and that there is always potential for misinformation to arise.

\section{FURTHER WORK}
Overall, this study furthers the understanding of false news measurement from different writing styles of journalists by analyzing its degree of falsity. Given the ongoing disagreement in conceptualizing false news and judgement of misinformation, further analysis of the degree of falsity is imperative. Although the extent to which an article becomes false is still vague, future studies must focus on partially false news which is more harmful. In the future, auxiliary information such as the headline, pictures and URLs can be used comprehensively to understand the characteristics of false news. The part of speech (POS) taggers such as NNG (common noun), NNP (proper noun), NP (pronoun) could be used to analyze the writing styles. Otherwise, different topics in more diversified categories, such as sports, politics, education, or economy, can be used in the experiment. Given that false news in other languages, like English, shows some similarities in linguistic characteristics, this measurement can be explored more extensively using other languages. For the purpose of comparison, better experiment is conducted with aligned texts of both real and false news. To generalize the scope of the work, a dataset that identify misinformation in pack journalistic traits would be necessary. 

More studies can be done to evaluate misinformation to establish verification practices for both journalists and readers. This is important because journalists are namely the credible contributors of information whose false news products can be pernicious. Given that pernicious false news in reality exists in the form of misinformation with partial falsity rather than complete falsity, future studies should delve into misinformation with more detailed measures. It can investigate how overtime different news outlets copy false information written by journalists and its impacts on other information sources. Finally, this work of measuring falseness hopefully instigates novel methods to tackle malicious false news in the future.

\bibliographystyle{unsrt}
\bibliography{references}


\end{document}